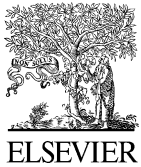

8th International Topical Meeting on Neutron Radiography, ITMNR-8, 4-8 September 2016, Beijing, China

# A neutron detector with spatial resolution of submicron using fine-grained nuclear emulsion

N. Naganawa[a]*, S. Awano[b], M. Hino[c], M. Hirose[d], K. Hirota[b], H. Kawahara[b], M. Kitaguchi[b], K. Mishima[e], T. Nagae[d], H. M. Shimizu[b], S. Tasaki[f], A. Umemoto[b].

[a]*Institute of Materials and Systems for Sustainability, Nagoya University, Chikusa, Nagoya, 464-8602, Japan*
[b]*Department of Physics, Nagoya University, Chikusa, Nagoya, 464-8602, Japan*
[c]*Research Reactor Institute, Kyoto University, Kumatori, Osaka 590-0494, Japan*
[d]*Department of Physics, Kyoto University, Kyoto 606-8502, Japan*
[e]*High Energy Accelerator Research Organization, Tokai, Ibaraki 319-1106, Japan*
[f]*Department of Nuclear Engineering, Kyoto University, Kyoto 615-8540, Japan*

**Abstract**

We have been developing a neutron detector with spatial resolution of submicron by loading $^6$Li into fine-grained nuclear emulsion. By exposure to thermal neutrons, tracks from neutron capture events were observed. From their grain density, spatial resolution was estimated. Detection efficiency was also measured by an experiment with cold neutrons.

## 1. Introduction

Recently, importance of neutron detectors with high spatial resolution is increasing in neutron imaging. We have been developing detectors with spatial resolution of submicron by loading $^6$Li, a nuclide with large absorption cross section to neutron, into fine-grained nuclear emulsion, a 3-dimensional tracking detector with high spatial resolution. With the detector, resolution of imaging can be better by more than an order of magnitude. Such emulsion detector is available to be applied for fundamental physics experiments with measurement of interference pattern of neutron, for example, searches for displacement from inverse square law in the position distribution of quantized states of ultra-cold neutrons under the earth's gravitational field (Nesvizhevsky et al. (2002), Abele et al. (2009), Ichikawa et al. (2014)).

* Corresponding author. Tel.: +81-52-789-3532; fax: +81-52-789-2864.
  E-mail address: naganawa@flab.phys.nagoya-u.ac.jp

## 2. A neutron detector using nuclear emulsion

### 2.1. Nuclear emulsion

Nuclear emulsion is a 3-dimensional tracking detector of charged particles with spatial resolution of submicron. It is a kind of photographic film which is a substrate made of plastic or glass coated with nuclear emulsion gel containing silver halide crystals dispersed in gelatin. Typical diameter of the crystals is about 200 nm. Their small size makes its spatial resolution high. When a charged particle passes through an emulsion layer, penetrating crystals along the passage, latent images which are clusters of silver atoms are formed inside the crystals. During development, penetrated crystals with latent images turn into silver grains with diameter of several times larger than the crystals, which are able to be observed by an optical microscope. Thus, a track of the particle is formed.

In 2010, a nuclear emulsion with higher resolution was developed at Nagoya University. It is a fine-grained nuclear emulsion with crystal size of 40 nm developed for detection of recoil tracks of WIMPs (Weakly Interacting Massive Particles) as a direct search of dark matter (Naka and Natsume (2007), Naka et al. (2013)).

### 2.2 Application to neutron detection with high spatial resolution

#### 2.2.1 Nuclear emulsion gel

The fine-grained nuclear emulsion gel is applied for our high spatial resolution neutron detector. The first reason is that it has high spatial resolution due to the small size of silver halide crystals. The second is that it has high rejection power to gamma ray background due to its low sensitivity. When used without sensitization, electron tracks from conversion of gamma ray are not detected.

#### 2.2.2 Detection principle

Nuclear emulsion will be a neutron detector by loading nuclides with large absorption cross section to neutron which emit charged particles after absorption. In the case of $^6$Li being loaded, neutrons will be absorbed by $^6$Li and converted into charged particles with the process shown in a formula (1) below.

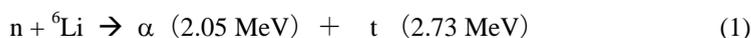

$$n + {}^6Li \rightarrow \alpha \ (2.05 \text{ MeV}) + t \ (2.73 \text{ MeV}) \quad (1)$$

When a neutron is absorbed by $^6$Li, an alpha particle and a triton with kinetic energy of 2.05 MeV and 2.73 MeV, respectively, are emitted. After development, a back-to-back topology formed by their tracks, as shown in Fig.1, will be observed under an optical microscope.

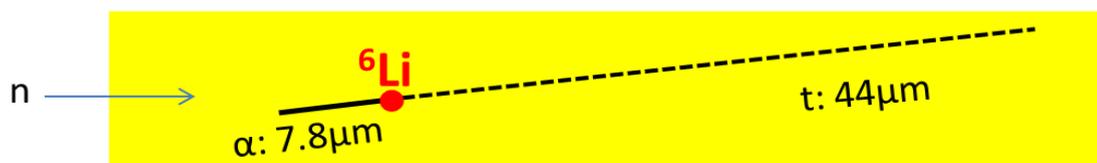

Fig. 1. After the absorption of neutron by $^6$Li loaded in nuclear emulsion, an alpha particle and a triton are emitted. A back-to-back topology is formed by their tracks. In the case of our fine-grained nuclear emulsion with LiNO$_3$ added, length of tracks of the alpha particle and the triton are 7.8 μm and 44 μm respectively from calculation by SRIM.

In a previous work (Taylor and Goldhaber (1935)), they loaded $^6$Li in nuclear emulsion and observed tracks of emitted particles after absorption of thermal neutrons.

In our study, we loaded $^6$Li into fine-grained nuclear emulsion. Grains constituting tracks were separately observed with its high spatial resolution. This enabled measurement of grain density (GD) of tracks which is number density of aligned silver grains along tracks. From measured GD and range of tracks, particles which created those tracks were identified. As a result, starting point of tracks of alpha particles and tritons, i.e. absorption points of neutrons, were possible to be defined.

In this study we loaded $^6$Li up to $1.2\times10^{-3}$ mol/cm$^3$ into nuclear emulsion, by adding LiNO$_3$ up to $1.6\times10^{-2}$ mol/cm$^3$. In this case, ranges of tracks of alpha particle and triton are 7.7 μm and 44 μm respectively from calculation by SRIM.

### 2.3 Fabrication of the detector

#### 2.3.1 Loading $^6$Li into nuclear emulsion gel

After fine-grained nuclear emulsion gel is melted at the temperature of 40 degrees centigrade, and while being mixed, aqueous solution of LiNO$_3$ is added by a micropipette. Thus, $^6$Li is added up to the amount corresponding to $1.2\times10^{-3}$ mol/cm$^3$ in dried emulsion gel. This amount is close to the maximum to be added. If added more, the gel will not be dried well and salt of LiNO$_3$ will be deposited on the surface of emulsion layer.

#### 2.3.2 Coating

For experiment in section 3.1, small amount of melted emulsion gel with LiNO$_3$ mixed was dropped on a 1 mm-thick glass plate by a micropipette, and spread by the tip of the pipette. For experiment in section 3.2, dip coating was done. After coating, emulsion samples were dried in the room air. After dry, thickness of emulsion layer was 27 μm.

#### 2.3.3 Packing of the detector

Nuclear emulsion is photosensitive. So, in order to protect our detector from light, we packed it in a light-tight package before exposure. The package is made of a foil consists of nylon (15 μm), polyethylene (13 μm), aluminum (7 μm), polyethylene (13 μm), and black polyethylene (80 μm). Thickness of each material is written in brackets.

## 3. Detection of neutron

### 3.1 Tracks and spatial resolution

To confirm the detection principle, a fabricated sample was exposed to thermal neutrons at CN-3 beam line of a reactor at Kyoto University Research Reactor Institute. After development, absorption events were observed under an optical microscope with an epi-illumination system. A microscopic image of one of the events observed is shown in Fig. 2. Tracks of an alpha particle and a triton which emerged from an absorption point forming back-to-back topology were observed. Those tracks were identified by GD. First half parts of alpha particles' tracks from emerging points had average GD of 1.4±0.4 grains/μm. Those of tritons' had that of 0.37±0.08 grains/μm. Average dE/dx of those particles for the same parts of tracks were calculated by SRIM to be $9.76\times10^2$ MeV/(g/cm$^2$) and $1.72\times10^2$ MeV/(g/cm$^2$) for alpha particles and tritons, respectively. Difference between the ratio of GD and dE/dx was not significant.

From the GD of an alpha particle, 1.4±0.4 grains/μm, the average distance between grains was 0.71 μm. From Poisson probability, alpha particle's emerging point exists at the distance between 0 to x μm upstream from the first grain with probability of 1-exp (- x/0.71). From this, the emerging point exists at (0.41±0.41) μm upstream of the first grain with the probability of 68 %. Thus, we estimated the spatial resolution of our detector for the detection of absorption points to be 0.4 μm. By raising GD through improvement of development method, higher resolution will be realized.

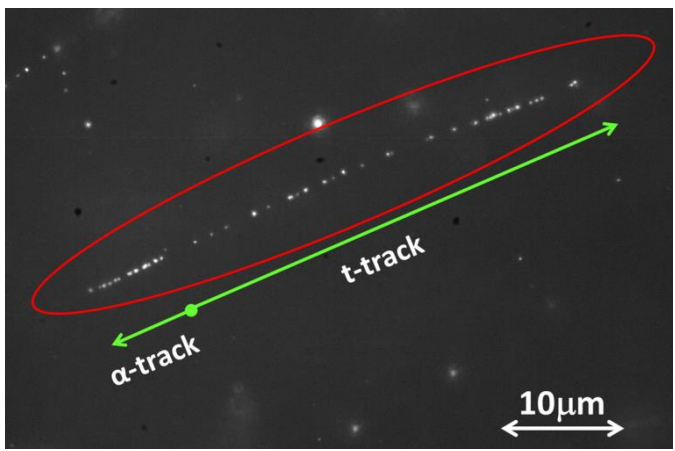

Fig. 2. An example of a neutron absorption event by $^6$Li observed under an optical microscope with an epi-illumination system (surrounded by a red ellipse). A pair of tracks of an alpha particle and a triton in back-to-back topology was recognized.

### 3.2 Detection efficiency

#### 3.2.1 Exposure

Detection efficiency was measured by exposing an emulsion detector to pulsed cold neutron with wavelength from 0.2 to 1.0 nm at Low divergence beam branch of BL05 in MLF/J-PARC (Mishima et al.(2009), Mishima (2015)). Repetition rate of the neutron pulse is 25 Hz. Experimental set up during the exposure is shown in Fig. 3. A pinhole with diameter of 1.1 mm made of a 1 mm-thick Cd plate was set at the upstream of an emulsion sample. At the downstream of the emulsion detector, intensity of neutron was measured by 0.97 MPa-$^3$He proportional counter, RS-P4-0812-223, of 25.4 mm diameter with a 0.5 mm-thick stainless steel wall. Wavelength distribution of neutrons which passed through the pinhole was as shown in Fig. 4. It was derived from time of flight information taken by the $^3$He counter, without setting the emulsion detector. The sample was exposed to (4.2±0.4)×10$^5$ neutrons in total which passed though the pinhole. The uncertainty of the neutron intensity is 10 %, mainly from that of detection efficiency of the $^3$He detector, from a conservative estimation.

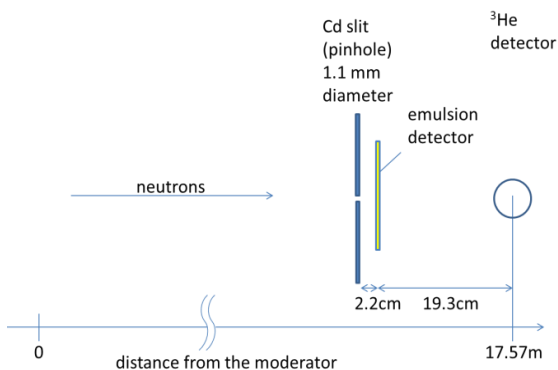
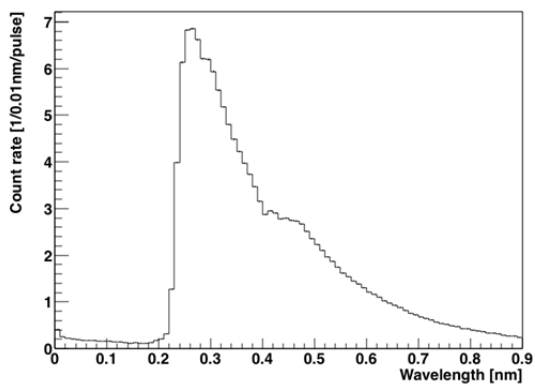

Fig. 3. The experimental set up during the exposure. At 2.2 cm upstream of the emulsion detector, a pinhole made of Cd with diameter of 1.1 mm was set. At the downstream a $^3$He detector was set.

Fig. 4. The wavelength distribution from the $^3$He detector set at the downstream.

*3.2.2 Analysis*

Emulsion detector was developed after the exposure. In order to calculate detection efficiency, number of absorption events in the area exposed to neutrons was necessary to be counted. It was done under the optical microscope with an epi-illumination system by human eyes. To minimize the time for counting, only sampled views were searched for events. As shown in Fig. 5, 25 sampled views on the emulsion, in and around the area of 1.1 mm diameter which corresponds to the downstream of the pinhole, was searched. Field of view of the microscope was (110 μm)$^2$. The whole thickness of the emulsion layer, 27 μm was searched. For the calculation of detection efficiency, total number of the events in 21 views inside the area of 1.1 mm diameter was used. Total area of 21 views was $2.5\times10^{-3}$ cm$^2$.

To determine the absorption point by the difference of GD of alpha and triton tracks, we only accept events with alpha and triton tracks which have at least three grains in the emulsion layer when they go out of the emulsion layer. This requirement reduces the efficiency of counting neutrons to (81±9) % of absorbed neutron, which is calculated numerically from geometry.

*3.2.3 Results*

37 absorption events were detected in the 21 views with the area of $2.5\times10^{-3}$ cm$^2$. On the other hand, the number of exposed neutrons was $(1.1\pm0.1)\times10^5$ in the same area. By considering the wavelength distribution in Fig. 4 and absorption cross section being proportional to inverse of velocity, expected number of absorption events to be occurred in the same area was calculated to be (44±5) events. And due to the requirement to tracks discussed in the previous section, (81±9) % of them, (36±6) events were expected to be detected. The number of the detected events was consistent with that of the expected.

Detection efficiency calculated from this result is $(3.3\pm0.6)\times10^{-4}$ for the cold neutrons in this experiment. From the extrapolation of the result, detection efficiency for thermal neutrons with velocity of 2200 m/s is $(1.5\pm0.5)\times10^{-4}$.

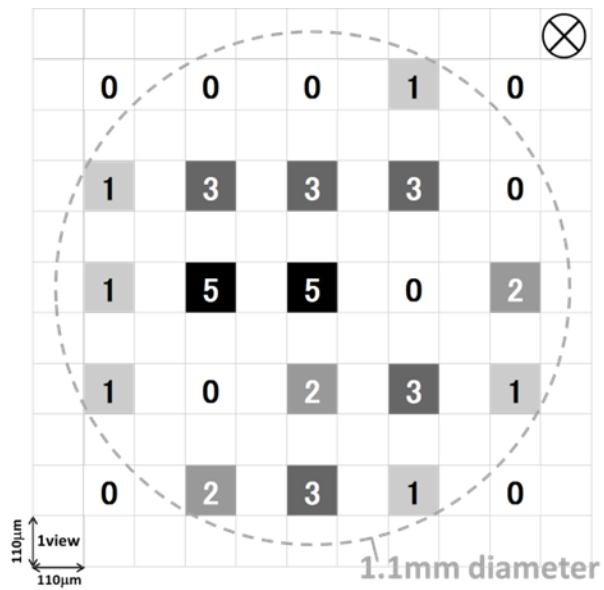

Fig. 5. Events are searched in and around the circled area of 1.1 mm diameter every other view. Each square corresponds to the field of view of the microscope which is (110 μm)$^2$. Views searched for events are shown with numbers of counted events. Total number of events in 21 views inside the area of 1.1 mm diameter was used for the measurement of detection efficiency.

## 4. Conclusion and outlook

We have been developing a neutron detector with spatial resolution of submicron by loading $^6$Li into fine-grained nuclear emulsion. Events of neutron absorption by $^6$Li were successfully observed from the exposure to thermal neutron. From the grain density (GD) of observed alpha tracks, spatial resolution of detecting absorption point was estimated to be 0.4 μm, Further improvement of the resolution is possible with improvement of development method which raises GD.

Detection efficiency was measured by the test exposure to cold neutrons. For exposed neutrons, it turned out to be $(3.3\pm0.6)\times10^{-4}$, which was consistent with expectation. From extrapolation of this result, it turned out to be $(1.5\pm0.5)\times10^{-4}$ to thermal neutrons with velocity of 2200 m/s.

With this detector, the spatial resolution of neutron radiography is possible to be higher by more than an order of magnitude, if with an ideal neutron exposure. The detector can also be used for fundamental physics experiments with measurement of interference pattern of neutron. We will study the uniformity of the detection efficiency and distortion of the detector.

Besides this detector, we are also developing a detector with a very thin layer including $^{10}B$ coated with nuclear emulsion for higher resolution less than 100 nm.

## 5. Acknowledgments

We acknowledge Dr. Y. Seki for sharing his beam time and helping our exposure at KURRI. The efficiency measurement was approved by the Neutron Science Proposal Review Committee of J-PARC/MLF (Proposal No. 2014B0270 and 2015A0242) and supported by the Inter-University Research Program on Neutron Scattering of IMSS, KEK. This work was supported by JSPS KAKENHI Grant Number JP26800132.